\documentclass[iop]{emulateapj}

\shorttitle{Dark matter fractions in three lens galaxies}
\shortauthors{Bate et al.}

\begin{document}

\title{A microlensing measurement of dark matter fractions in three lensing galaxies\altaffilmark{1}}

\author{N. F. Bate, D. J. E. Floyd\altaffilmark{2}, R. L. Webster and J. S. B. Wyithe}
\affil{School of Physics, The University of Melbourne, Parkville, Vic, 3010, Australia}
\email{nbate@physics.unimelb.edu.au}

\altaffiltext{1}{This paper uses data gathered with the 6.5 meter Magellan Telescopes located at Las Campanas Observatory, Chile.}
\altaffiltext{2}{OCIW, Las Campanas Observatory, Casilla 601, Colina El Pino, La Serena, Chile}

\begin{abstract}
Direct measurements of dark matter distributions in galaxies are currently only possible through the use of gravitational lensing observations. Combinations of lens modelling and stellar velocity dispersion measurements provide the best constraints on dark matter distributions in individual galaxies, however they can be quite complex. In this paper, we use observations and simulations of gravitational microlensing to measure the smooth (dark) matter mass fraction at the position of lensed images in three lens galaxies: \objectname{MG 0414+0534}, \objectname{SDSS J0924+0219} and \objectname{Q2237+0305}. The first two systems consist of early-type lens galaxies, and both display a flux ratio anomaly in their close image pair. Anomalies such as these suggest a high smooth matter percentage is likely, and indeed we prefer $\sim50$ per cent smooth matter in MG 0414+0534, and $\sim80$ per cent in SDSS J0924+0219 at the projected locations of the lensed images. Q2237+0305 differs somewhat in that its lensed images lie in the central kiloparsec of the barred spiral lens galaxy, where we expect stars to dominate the mass distribution. In this system, we find a smooth matter percentage that is consistent with zero.
\end{abstract}

\keywords{dark matter --- gravitational lensing --- quasars: individual: MG 0414+0534 --- quasars: individual: SDSS J0924+0219 --- quasars: individual: Q2237+0305.}

\section{Introduction}

Gravitational lensing measurements provide the only direct method to probe the non-luminous matter component of lensing systems. Weak lensing measurements of cluster masses and mass distributions are becoming routine, however the measurement of the dark matter components of individual galaxies is a new and exciting prospect. By analysing gravitational microlensing signals in some strongly lensed quasars, we are able to infer the ratio of clumpy to continuously distributed matter along the line of sight to the background source. This gives us a probe of the dark matter content in a lensing galaxy at projected radii of $\sim2$--10kpc from the centre of the galaxy (\citealt{sw04}; \citealt{pooley+09}).

Recent years have seen considerable advancement in the mapping of dark and stellar mass in lensing galaxies (e.g. \citealt*{keeton+98}; \citealt*{ferreras+05}; \citealt{barnabe+09}). Usually, the total mass within the Einstein Radius is constrained by modelling the lensing galaxy to fit the observed lensed image positions. Photometry of the lensing galaxy, in combination with stellar population synthesis models, provides the distribution of stellar mass. Finally, observations of stellar velocity dispersions can be used to break the mass-sheet and mass-anisotropy degeneracies, and thus constrain the overall mass density profile (see for example \citealt{kt03}; \citealt{tk04}; \citealt{ferreras+05}; \citealt*{ferreras+08}; \citealt{barnabe+09}; \citealt{auger+09}, and references therein).

These analyses can provide a very detailed picture of the structure of lensing galaxies. However, they are quite complex, relying on often difficult observations and detailed modelling. A complementary method exists, in which observations of microlensing in the lensed quasar images are used to constrain the dark matter percentage along those lines of sight directly.

Cosmological microlensing occurs when the light path to a lensed quasar image intersects a starfield in a foreground lensing galaxy. The lensing galaxy as a whole magnifies the lensed image; microlensing by individual stars induces variations about this macro-magnification. This is most readily detected in lightcurves of quasar images, where relative motion between observer, lens and source causes uncorrelated fluctuations in brightness between lensed images. This effect was first observed in the lensed quasar Q2237+0305 \citep{irwin+89}.

In some lensing systems we observe a close pair of quasar images. Basic lensing theory suggests that these two images should have approximately equal magnifications (\citealt{cr79}; \citealt{bn86}). On the contrary, in eight out of ten known cases we find that the brightness of the image located at the saddle point in the time delay surface is suppressed relative to the image at the minimum in the time delay surface \citep{pooley+07}. Microlensing is one possible explanation for such flux ratio anomalies. However, microlensing by a purely stellar component is not sufficient. \citet{sw02} showed that this discrepancy could be accounted for by adding a significant smooth matter component to the lens at the image positions, since minimum and saddle point images are microlensed differently when a smooth matter component is added.

We have previously developed a technique for using single-epoch multi-wavelength observations of anomalous lensed quasars to constrain the radius and radial profile of the background quasar accretion discs (\citealt{bfww08}; \citealt{fbw09}). In those analyses, we marginalised over the smooth matter percentage in the lens as a nuisance parameter. Here, we turn the problem around and instead marginalise over the quasar parameters to obtain constraints on the smooth matter percentage in the lens at the image positions.

Rough microlensing measurements of smooth matter percentages have been reported previously. Spectroscopy of SDSS J0924+0219 undertaken by \citet{keeton+06} suggested a smooth matter percentage of 80 to 85 per cent in that lens at the location of the $D$ and $A$ images. Using X-ray monitoring of HE 1104+1805, \citet{chartas+09} reported a smooth matter percentage of $\sim80$ per cent is favoured. \citet{pooley+09} measured the smooth matter percentage in PG 1115+080 to be $\sim90$ per cent, using X-ray observations. \citet{metal08} found a weak trend supporting this result. Most recently, \citet{dai+09} favoured a smooth matter fraction of $\sim70$ per cent using X-ray and optical monitoring of RXJ 1131-1231. Microlensing analyses consistently predict a significant smooth matter percentage in the lensing galaxy at the position of anomalous images.

In this paper, we present constraints on the dark matter percentages in three lensing galaxies: MG 0414+0534, SDSS J0924+0219 and Q2237+0305. MG 0414+0534 and SDSS J0924+0219 are both lensed by early-type galaxies, and consist of close image pairs displaying a flux ratio anomaly. MG 0414+0534 is moderately anomalous, whereas SDSS J0924+0219 is the most anomalous lensed quasar currently known. Q2237+0305 differs from the previous sources in two key ways: it is lensed by a barred spiral galaxy, and it does not contain a close image pair. Nevertheless, it is known to be affected by microlensing (e.g. \citealt{irwin+89}).

This paper is laid out as follows: in Section \ref{sec:obs} we discuss the observational data on the three systems of interest. The simulation technique is briefly described in Section \ref{sec:sims}. We present our results and discussion in Section \ref{sec:results}, and conclude in Section \ref{sec:conclusions}. Throughout this paper we use a cosmology with $H_0=70\rm{kms^{-1}Mpc^{-1}}$, $\Omega_m=0.3$ and $\Omega_{\Lambda}=0.7$.

\section{Observational data}
\label{sec:obs}

\subsection{MG 0414+0534}
MG 0414+0534 was discovered by \citet{hewitt+92}. It consists of a background quasar at $z_s=2.64$ \citep{lejt95} and a foreground early-type lensing galaxy at $z_l=0.96$ \citep{tk99}. Four images of the quasar are observed, with the close image pair (images $A_1$ and $A_2$) displaying a flux ratio anomaly. This anomaly is weak in both the mid-infrared ($A_2/A_1 = 0.90 \pm 0.04$ on 2005 October 10, \citealt{minezaki+09}) and the radio ($A_2/A_1 = 0.90 \pm 0.02$ on 1990 April 2, \citealt{kh93}), but somewhat stronger in the optical ($A_2/A_1 = 0.45\pm 0.06$ on 1991 November 2-4, \citealt{sm93}).

In our analysis, we used three epochs of multi-wavelength MG 0414+0534 observations, presented in Table \ref{0414obs}. The first two epochs were archival HST data, obtained from the CASTLES Survey webpage\footnote{http://cfa-www.harvard.edu/castles/} \citep{fls97}. The third epoch was obtained by us using the Magellan 6.5-metre Baade telescope. These data were first presented in \citet{bfww08}.

\begin{deluxetable}{lrll}
\tablecaption{Observed flux ratios in MG 0414+0534\label{0414obs}}
\tablehead{
\colhead{Filter} & \colhead{$\lambda_c$ (\AA)} & \colhead{$F_{obs} = A_2/A_1$} & \colhead{Date}}
\startdata
    $H$ & 16500 & $0.67\pm0.05$ & 2007 November 3\\
    $J$ & 12500 & $0.60\pm0.2$ & 2007 November 3\\
    $z^\prime$ & 9134 & $0.34\pm0.1$ & 2007 November 3\\
    $i^\prime$ & 7625 & $0.26\pm0.1$ & 2007 November 3\\
    $r^\prime$ & 6231 & $0.21\pm0.1$ & 2007 November 3\\
    F205W & 20650 & $0.83\pm0.03$ & 1997 August 14 \\
    F110W & 11250 & $0.64\pm0.04$ & 1997 August 14 \\
    F814W & 7940 & $0.47\pm0.01$ & 1994 November 8 \\
    F675W & 6714 & $0.40\pm0.01$ & 1994 November 8 \\
\enddata
\tablecomments{Central wavelengths $\lambda_c$ and observed (anomalous) flux ratios $F_{obs}$ between images $A_2$ and $A_1$ in MG 0414+0534, in each of nine filters. The 2007 November 3 observations were taken with the IMACS and PANIC instruments on the Magellan 6.5-m Baade telescope \citep{bfww08}. The 1997 August 14 observations were taken with the NICMOS instrument on \textit{HST} (obtained from the CASTLES Survey web page. The 1994 November 8 observations were taken with the WFPC2 instrument on \textit{HST} (\citealt{fls97}).}
\end{deluxetable}

\subsection{SDSS J0924+0219}

SDSS J0924+0219 is the most anomalous lensed quasar currently known. The minimum image $A$ has been observed to be a factor of $\sim20$ brighter than the saddle point image $D$ in the optical \citep{keeton+06}. The quasar was discovered by \citet{inada03} in Sloan Digital Sky Survey (SDSS) imaging, and consists of an early-type lensing galaxy at $z_l = 0.394$ \citep{eigenbrod06a} and a background quasar at $z_s=1.524$ \citep{inada03}.

Again, we use three epochs of observational data. These are presented in Table \ref{0924obs}. The 2008 March 21 data were obtained by us using the Magellan 6.5-metre Baade telescope \citep{fbw09}. The 2003 November 18-23 data were taken using the HST/NICMOS and WFPC2 instruments as part of the CASTLES Survey \citep{keeton+06}. The 2001 December 15 were obtained by \citet{inada03} using the MagIC instrument on the Baade telescope, and re-reduced by us (details can be found in \citealt{fbw09}).

\begin{deluxetable}{lrll}
\tablecaption{Observed flux ratios in SDSS J0924+0219\label{0924obs}}
\tablehead{
\colhead{Filter} & \colhead{$\lambda_c$ (\AA)} & \colhead{$F_{obs}={D}/{A}$} & \colhead{Date}}
\startdata
$H$ & $16500\pm1450$ & $0.23\pm0.05$ & 2008 March 21 \\
$J$ & $12500\pm800$ & $0.15\pm0.05$ & 2008 March 21 \\
$Y$ & $10200\pm500$ & $0.14\pm0.05$ & 2008 March 21 \\
$z^\prime$ & $9134\pm800$ & $0.19\pm0.10$ & 2008 March 21 \\
$i^\prime$ & $7625\pm650$ & $0.16\pm0.10$ & 2008 March 21 \\
$r^\prime$ & $6231\pm650$ & $0.10\pm0.10$ & 2008 March 21 \\
$g^\prime$ & $4750\pm750$ & $0.08\pm0.08$ & 2008 March 21 \\
$i^\prime$ & $7625\pm650$ & $0.08\pm0.05$ & 2001 December 15 \\
$r^\prime$ & $6231\pm650$ & $0.07\pm0.05$ & 2001 December 15 \\
$g^\prime$ & $4750\pm750$ & $0.06\pm0.05$ & 2001 December 15 \\
$u^\prime$ & $3540\pm310$ & $<$0.09 & 2001 December 15 \\
F160W & $15950\pm2000$ & $0.08\pm0.01$ & 2003 November 18 \\
F814W & $8269\pm850$ & $0.05\pm0.005$ & 2003 November 18-19 \\
F555W & $5202\pm600$ & $0.05\pm0.01$ & 2003 November 23 \\
\enddata
\tablecomments{Central wavelengths $\lambda_c$ and observed (anomalous) flux ratios $F_{obs}$ between images $A$ and $D$ in SDSS J0924+0219, in each filter. The 2008 March 21 observations were taken with the IMACS and PANIC instruments on the Magellan 6.5-m Baade telescope \citep{fbw09}. The 2001 December 15 data were taken using MagIC on Baade \citep{inada03}. The 2003 data were taken using HST/NICMOS and WFPC2 for the CASTLES Survey \citep{keeton+06}.}
\end{deluxetable}

\subsection{Q2237+0305}

Q2237+0305 is perhaps the most well-studied gravitationally lensed quasar. It was discovered by \citet{huchra+85}, and consists of a lensing galaxy at $z_l=0.0394$ and a background quasar at $z_s=1.695$. The two previous sources had early type lensing galaxies; the lens in Q2237+0305 is a barred spiral. Near-perfect alignment between observer, lens and quasar results in four virtually symmetric images of the background source, located in the bulge of the lensing galaxy. The optical depth to stars is therefore quite high, making the system an excellent target for microlensing analyses. Typically, this is taken to mean the smooth matter percentage in microlensing simulations can be set to zero. We will test this assumption here. 

Q2237+0305 also differs from the two previous sources in that it does not contain a close image pair displaying  a flux ratio anomaly. Nevertheless, its images are known to vary in brightness due to microlensing. We choose to examine images $A$ and $B$ both because they are well modelled, and they are roughly equi-distant from the centre of the lensing galaxy. 

Our observational data were obtained from \citet{eigenbrod+08b}, in which 43 epochs of Q2237+0305 spectroscopic data from the FORS1 spectrograph on the Very Large Telescope (VLT) were analysed. Eigenbrod and collaborators deconvolved their Q2237+0305 spectra into a broad emission line component, a continuum emission component, and an iron pseudo-continuum. The continuum emission was fit with a power-law of the form $f_\nu \propto \nu^{\alpha_\nu}$. The power-law fit was then split into six wavelength bands, each with a width of 250\AA~in the quasar rest frame, and integrated in each band. The result is pseudo-broadband photometry in six wavebands, with contamination from broad emission lines and the iron continuum removed. We selected two epochs from these data for our analysis, separated by approximately a year. The flux ratios are presented in Table \ref{2237obs}. Following the \citet{eigenbrod+08b} numbering, the 2005 November 11 dataset is epoch number 17, and the 2006 November 10 dataset is epoch number 28. 

\begin{deluxetable*}{lrrll}
\tablecaption{Observed flux ratios in Q2237+0305\label{2237obs}}
\tablehead{
\colhead{Band} & \colhead{Emitted $\lambda_c$ (\AA)} & \colhead{Observed $\lambda_c$ (\AA)} & \colhead{$B/A$} & \colhead{Date}}
\startdata
    1 & $1625\pm125$ & $4379\pm337$ & $0.52\pm0.02$ & 2005 November 11 \\
    2 & $1875\pm125$ & $5053\pm337$ & $0.51\pm0.02$ & 2005 November 11 \\
    3 & $2125\pm125$ & $5727\pm337$ & $0.50\pm0.01$ & 2005 November 11 \\
    4 & $2375\pm125$ & $6401\pm337$ & $0.49\pm0.01$ & 2005 November 11 \\
    5 & $2625\pm125$ & $7074\pm337$ & $0.48\pm0.01$ & 2005 November 11 \\
    6 & $2875\pm125$ & $7748\pm337$ & $0.47\pm0.01$ & 2005 November 11 \\
    1 & $1625\pm125$ & $4379\pm337$ & $0.33\pm0.02$ & 2006 November 10 \\
    2 & $1875\pm125$ & $5053\pm337$ & $0.34\pm0.02$ & 2006 November 10 \\
    3 & $2125\pm125$ & $5725\pm337$ & $0.35\pm0.01$ & 2006 November 10 \\
    4 & $2375\pm125$ & $6401\pm337$ & $0.36\pm0.02$ & 2006 November 10 \\
    5 & $2625\pm125$ & $7074\pm337$ & $0.37\pm0.02$ & 2006 November 10 \\
    6 & $2875\pm125$ & $7748\pm337$ & $0.37\pm0.01$ & 2006 November 10 \\
\enddata
\tablecomments{Two epochs of observational $B/A$ flux ratios for Q2237+0305, obtained from \citet{eigenbrod+08b}. Photometry was extracted from spectra obtained with the FORS1 spectrograph on the Very Large Telescope (VLT) at the European Southern Observatory (ESO). Filter wavelengths are in the rest frame of the source quasar, located at a redshift of $z_s=1.695$. Following the \citet{eigenbrod+08b} numbering, the 2005 November 11 dataset is epoch number 17, and the 2006 November 10 dataset is epoch number 28.}
\end{deluxetable*}



\section{Microlensing simulations}
\label{sec:sims}

The simulation technique used in this work has been presented previously in \citet{bfww08} and \citet*{fbw09}. In those papers, we marginalised over the smooth matter percentage $s$ as a nuisance parameter. Here, we instead marginalise over the accretion disc radius $\sigma_0$ and the power-law index $\zeta$ relating the radius of the accretion disc to the observed wavelength.

Our microlensing simulations were conducted using an inverse ray-shooting technique (\citealt{krs86}; \citealt{wpk90}). The key parameters in these simulations are the convergence $\kappa_{tot}$, which is a measure of the focussing power of the lens, and shear $\gamma$, which is a measure of the distortion introduced  by the lens. The lensing parameters used in this analysis are presented in Table \ref{tab:lensparams}.

\begin{deluxetable}{lcccc}
\tablecaption{Lensing parameters for the images of interest in this analysis\label{tab:lensparams}}
\tablehead{
\colhead{Quasar} & \colhead{Image} & \colhead{$\kappa_{tot}$} & \colhead{$\gamma$} & \colhead{$\mu_{tot}$}}
\startdata
    MG 0414+0534 & $A_1$ & 0.472 & 0.488 & 24.2 \\
    MG 0414+0534 & $A_2$ & 0.485 & 0.550 & -26.8 \\
    SDSS J0924+0219 & $A$ & 0.502 & 0.458 & 26.2 \\
    SDSS J0924+0219 & $D$ & 0.476 & 0.565 & -22.4 \\
    Q2237+0305 & $A$ & 0.413 & 0.382 & 5.03 \\
    Q2237+0305 & $B$ & 0.410 & 0.384 & 4.98 \\
\enddata
\tablecomments{Convergence $\kappa_{tot}$, shear $\gamma$ and magnification $\mu_{tot}$ for each of the lensed images examined in this paper. A negative total magnification is interpreted as a parity flip. MG 0414+0534 parameters were obtained from \citet{wms95}. SDSS J0924+0219 parameters were obtained from \citet{keeton+06}. Q2237+0305 were obtained via private communication with Cathryn Trott, based on modelling in \citet{trott+09}.}
\end{deluxetable}

The convergence can be split into two components $\kappa_{tot} = \kappa_* + \kappa_s$, where $\kappa_*$ describes a clumpy stellar component and $\kappa_s$ a smoothly distributed component. We define the smooth matter percentage $s$ to be the ratio of the continuously distributed component to the total convergence:

\begin{equation}
s = \kappa_s / \kappa_{tot}
\end{equation}
We allowed the smooth matter percentage to vary from 0 to 99 per cent, in 10 per cent increments. The smooth matter percentage is thus relatively coarsely sampled; our simulations were optimised to probe the accretion disc sources in each system.

The microlenses in our simulations were drawn from a Salpeter mass function $dN/dM \propto M^{-2.35}$ with a mass range $M_{max}/M_{min} = 50$. Physical sizes are therefore scaled by the average Einstein Radius projected on to the source plane $\eta_0$, which varies from system to system. Magnification maps were generated covering an area of $24\eta_0 \times 24\eta_0$, with a resolution of $2048\times2048$ pixels. Ten maps were generated for each image and smooth matter percentage.

As discussed in \citet{bfww08}, we randomly selected source positions in each combination of magnification maps, to build up a simulated library of flux ratio curves as a function of wavelength. Comparing these with the observations allows us to construct a three dimensional likelihood distribution for the observed flux ratio spectrum $F^{obs}$ given three model parameters: the radius of the quasar source in the bluest filter $\sigma_0$, the power-law index relating observed wavelength to radius of the source $\zeta$, and the smooth matter percentage $s$. We can convert these likelihoods to an \textit{a posteriori} probability distribution for the three model parameters given the observations using Bayes theorem:

\begin{equation}
\frac{\rm{d}^3P}{\rm{d}\sigma_0\rm{d}\zeta\rm{d}s} \propto L(F^{obs}|\sigma_0,\zeta,s) \frac{\rm{d}P_{prior}}{\rm{d}\sigma_0} \frac{\rm{d}P_{prior}}{\rm{d}\zeta} \frac{\rm{d}P_{prior}}{\rm{d}s}
\end{equation}

Uniform priors were used for the two dimensionless quantities, smooth matter percentage $s$ and power-law index $\zeta$. A logarithmic prior was used for the radius $\sigma_0$. We note that this differs slightly from the analyses in \citet{bfww08} and \citet{fbw09}, where a uniform prior was also used for $\sigma_0$. We will briefly discuss prior dependence in Section \ref{sec:results}. We marginalise over the accretion disc parameters $\sigma_0$ and $\zeta$ to obtain a probability distribution for the smooth matter percentage $s$:

\begin{equation}
\frac{\rm{d}P}{\rm{d}s} = \int \int \frac{\rm{d}^3P}{\rm{d}\sigma_0\rm{d}\zeta\rm{d}s} \rm{d}\sigma_0 \rm{d}\zeta
\end{equation}

Our analysis focusses on two lensed images in each system. By dealing with flux ratios between images only, we remove the intrinsic quasar flux from the problem, provided the difference in light travel time between images is short. We assume that the smooth matter percentages in the two lensed images are identical. This is reasonable for the two anomalous systems, MG 0414+0534 and SDSS J0924+0219, as the anomalous images lie very close to each other. In Q2237+0305, where the images are widely separated, we choose to analyse image $A$ and $B$ only as they are essentially equi-distant from the centre of the lensing galaxy, and do not lie atop any obvious spiral features.

The probability distributions we obtain for smooth matter percentage are presented for MG 0414+0534 (Figure \ref{0414smooth}),  SDSS J0924+0219 (Figure \ref{0924smooth}) and Q2237+0305 (Figure \ref{2237smooth}). The dashed histograms show the differential probability distributions, and the solid lines show the cumulative probability distributions.

\section{Results and discussion}
\label{sec:results}

We obtain the following formal constraints on smooth matter percentage at the image positions in each system: $50^{+30}_{-40}$ per cent in MG 0414+0534, $80^{+10}_{-10}$ per cent in SDSS J0924+0219, and $\leq50$ in Q2237+0305 (68 per cent confidence limits are quoted). Our simulations are not optimised to probe smooth matter percentage; we sample smooth matter parameter space only sparsely, in order to reduce simulation time. These results should therefore be considered estimates, rather than exact measurements. Nevertheless, they provide an interesting, and currently poorly explored, measurement of smooth matter content within only a few effective radii of early type galaxies.

\begin{figure}
  \plotone{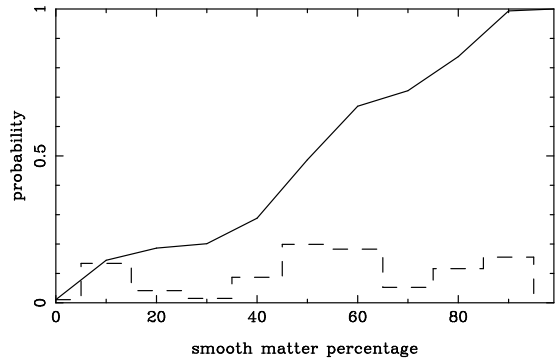}
  \caption{Probability distribution for smooth matter percentage in MG 0414+0534. The differential distribution (dashed line) and cumulative distribution (solid line) are both provided.\label{0414smooth}}
\end{figure}

\begin{figure}
  \plotone{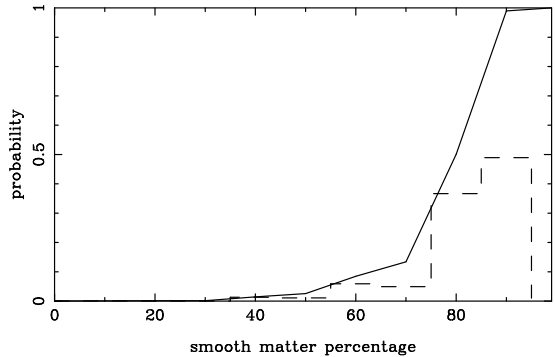}
  \caption{Probability distribution for smooth matter percentage in SDSS J0924+0219. The differential distribution (dashed line) and cumulative distribution (solid line) are both provided.\label{0924smooth}}
\end{figure}

In MG 0414+0534, the differential probability distribution (Figure \ref{0414smooth}, dashed line) does not particularly favour any single smooth matter percentage. This leads to a quite broad formal constraint on the smooth matter percentage.

Conversely, our measured smooth matter percentage in SDSS J0924+0219 is high, as we would expect for such an anomalous system. There is one other measurement of smooth matter percentage of this system in the literature. \citet{keeton+06} estimated it to be 80 to 85 per cent, based on the estimated size of the broad emission line region in the system. Our analysis is focussed on the accretion disc, and finds essentially identical results. We do use the \citet{keeton+06} observational data in obtaining our constraints, however excluding it and working only with the Magellan data does not significantly alter our results.

\begin{figure}
  \plotone{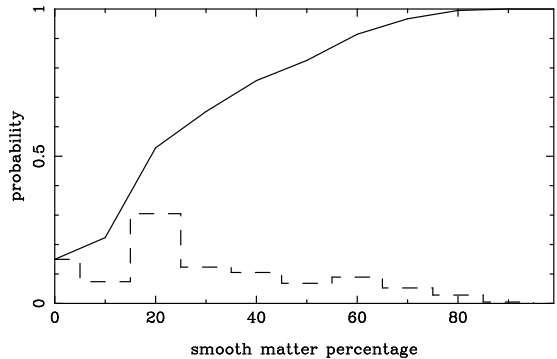}
  \caption{Probability distribution for smooth matter percentage in Q2237+0305 at the image $A$ and $B$ positions. The $B/A$ flux ratio was used as those images are at roughly the same projected distance from the centre of the lensing galaxy ($\sim0\farcs95$). The differential distribution (dashed line) and cumulative distribution (solid line) are both provided.\label{2237smooth}}
\end{figure}

As has been discussed earlier, the lensed images in Q2237+0305 lie in the bulge of the lensing galaxy. Stars are therefore expected to dominate the microlensing signal, rather than a smooth matter component. Indeed, we find a smooth matter percentage that is consistent with zero in this system. There is a peak in the differential probability distribution (Figure \ref{2237smooth}, dashed line) at $\sim20$ per cent. Though we are reluctant to suggest that this peak is real, we do note that such a feature could be evidence of additional absorbing features along the line of sight (see for example \citealt{foltz+92}, who reported MgII absorption features in the spectrum of Q2237+0305 at a redshift of 0.97).

To confirm that our results are not dominated by the choice of prior probability for the radius of the background quasar accretion disc $\sigma_0$, we repeated our analysis using a uniform prior rather than a logarithmic prior. Within their errors, we found no significant variation in our constraints on smooth matter percentage in any of our systems.

We can perform a simple calculation to obtain a rough theoretical prediction of the percentage of dark matter we expect to see at the image positions in each source. The method is briefly described by \citet{sw02}, but will be repeated here for clarity. It is only applicable to MG 0414+0534 and SDSS J0924+0219 as it assumes an early type lensing galaxy. We begin by working out a stellar surface mass density $\Sigma_s$ at the image positions. To do this, we take the observed effective radius of the lensing galaxy $R_e$, and compare it with Figure 10 in \citet{bernardi03} to obtain a $g$-band surface brightness in magnitudes per square arcsecond (assuming no evolution in early type galaxies between redshift $z=0$ and the redshift of the lensing galaxy). 

\citet{kauffmann03} provide a relationship between mass-to-light ratio $M/L$ and $g$-band magnitude derived from $10^5$ Sloan galaxies in their Figure 14. Using the $g$-band surface brightness obtained from \citet{bernardi03}, we can get a rough mass-to-light ratio for our lensing galaxy. We convert our surface brightness from magnitudes per square arcsecond into solar luminosities per square parsec using the following relationship:

\begin{equation}
  S[\rm{mag}/\rm{arcsec}^2] = M_{\odot g} + 21.572 -2.5\rm{log}_{10}S[\rm{L}_\odot/\rm{pc}^2]
\end{equation}
where S is the surface brightness and $M_{\odot g}=5.45$ is the solar magnitude in the $g$-band. Now that we have the lens galaxy surface brightness in solar magnitudes, we can use the mass-to-light ratio to convert into a stellar surface mass density in solar masses per square parsec. This is the stellar surface mass density at the effective radius $R_e$ -- we use the deVaucoulers profile to extrapolate this value out to the Einstein Radius, which is the approximate location of the lensed images.

The final piece of information we need is the surface mass density at the image positions. The critical surface mass density $\Sigma_{cr}$ for lensing is obtained from the following relationship:

\begin{equation}
  \Sigma_{cr} = \frac{c^2}{4\pi G} \frac{D_s}{D_d D_{ds}}
\end{equation}
where $D_d$ is the angular diameter distance to the lens, $D_s$ is the angular diameter distance to the source, and $D_{ds}$ is the angular diameter distance between lens and source. The other symbols have their usual meanings. For an isothermal sphere, the surface mass density at the Einstein Radius is half the critical surface mass density $\Sigma_{cr}$. The smooth matter percentage $s$ is simply obtained as follows:

\begin{equation}
  s = 1 - \frac{\Sigma_s}{0.5\Sigma{cr}}
\end{equation}

MG 0414+0534 has a deVaucoulers effective radius $R_e = 0\farcs78$ \citep{ketal00}, which gives a $g$-band magnitude of $\sim21.75$ and a mass-to-light ratio $M/L \sim4$. This gives a stellar surface mass density at the effective radius of $514 M_\odot/pc^2$. For an Einstein Radius of $1\farcs15$ \citep{trotter00}, the stellar surface mass density on the Einstein Ring is smaller by a factor of 0.46, giving $\Sigma_s=236 M_\odot/pc^2$. For a source at $z=2.64$ and a lens at $z=0.96$, the critical surface mass density for lensing is $\Sigma_{cr} = 2189 M_\odot/pc^2$. This gives a theoretically predicted smooth matter percentage of 78 per cent. Note that this number differs slightly from the figure quoted in \citet{sw02} as they were using preprints of \citet{bernardi03} and \citet{kauffmann03}. This prediction is on the high side of our measured smooth matter percentage of $50^{+30}_{-40}$ per cent, although it is formally consistent.

SDSS J0924+0219 has two measured effective radii in the literature: $R_e=0\farcs31\pm0\farcs02$ \citep{metal06} and $R_e=0\farcs50\pm0\farcs05$ \citep{eigenbrod06a}. Both results were obtained using {\it HST} data. \citet{metal06} fit a deVaucoulers profile to the lensing galaxy, whereas \citet{eigenbrod06a} chose a two-dimensional exponential disk. Since an exponential profile is shallower than a deVaucoulers profile, it would tend to give a larger effective radius. This may allow the two effective radii to be reconciled, however we will deal with each case separately. The system has a lensing galaxy at $z=0.39$ and a source at $z=1.524$, giving a critical surface mass density of $\Sigma_{cr} = 2323 M_\odot/pc^2$.

For an effective radius $R_e=0\farcs50$ we expect a $g$-band surface brightness of $\sim20.5\rm{mag}/\rm{arcsec}^2$, and a mass-to-light ratio $M/L=5.6$. At the effective radius, the stellar surface mass density is $2285 M_\odot/pc^2$. For an Einstein Radius of $0\farcs85$ \citep{inada03} and a deVaucoulers profile, the stellar surface mass density at the Einstein Radius is lower by a factor of 0.35, giving $\Sigma_s = 791 M_\odot/pc^2$. Thus, we predict a smooth matter percentage of 32 per cent. This is significantly lower than our $1\sigma$ measured value of $80^{+10}_{-10}$. Even at the 95 per cent level, our measured smooth matter percentage is not consistent with the prediction of this rough calculation.

The situation is somewhat different for an effective radius $R_e = 0\farcs31$. Here, an early type galaxy with no evolution between $z=0$ and $z=0.39$ should have a $g$-band surface brightness of $\sim19.75\rm{mag}/\rm{arcsec}^2$ and a mass-to-light ratio $M/L=5$. This gives a stellar surface mass density of $4052 M_\odot/pc^2$ at the effective radius. Again assuming a deVaucoulers profile, the stellar surface mass density is reduced by a factor of 0.13 at the Einstein Radius of $0\farcs85$, giving $539 M_\odot/pc^2$. We therefore expect 46 per cent of the mass to be in stars, and 54 per cent in smooth matter. This is consistent with our measured smooth matter percentage at the 95 per cent level ($80^{+10}_{-30}$).

\begin{figure}
  \plotone{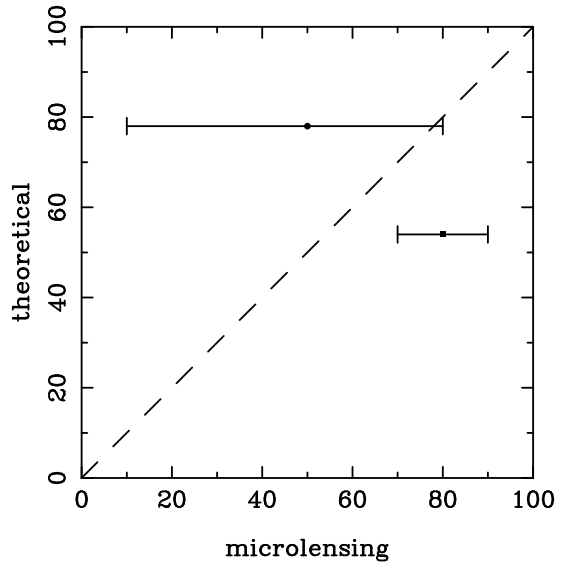}
  \caption{Comparison between smooth matter percentages obtained from microlensing measurements and rough theoretical calculations in this paper. No errors are given for the theoretical calculations (see Section \ref{sec:results}), although they should only be considered rough estimates. The dashed line shows 1:1 correspondence. Results for MG 0414+0534 (circle) and SDSS J0924+0219 (square) are shown, with 68 per cent confidence limits.\label{smooth_compare}}
\end{figure}

In Figure \ref{smooth_compare} we show a comparison between the results of this rough theoretical calculation and our microlensing measurements for MG 0414+0534 (circle) and SDSS J0924+0219 (square). No errors are provided for the rough theoretical calculations, although we emphasise that they should only be considered estimates. With only two data points, there does not seem to be any systematic difference between the microlensing measurements and the rough calculation.

It is, however, not correct to connect smooth matter percentage in our simulations directly with dark matter content in the lens. Intervening systems (which may not be detected otherwise) can contribute smooth matter to the surface mass density, as can gas and dust in the lens. As the lensing galaxies in both MG 0414+0534 and SDSS J0924+0219 are early type galaxies, we would expect this contribution to be very small. More importantly, \citet{lg06} showed that small compact masses can mimic a smooth matter component. This was first demonstrated analytically in \citet{dr81}.  For a source with a radius $0.1\eta_0$, \citet{lg06} find that compact masses smaller than $\sim0.01\rm{M}_\odot$ can mimic a smoothly distributed mass component. This may help to explain why our measured microlensing smooth matter percentage in SDSS J0924+0219 appears to be slightly higher than we would expect from a simple deVaucoulers galaxy profile in an isothermal dark matter profile.

We note that the three lensing systems analysed in this paper do not represent a fair statistical sample of gravitationally lensed quasars. The single-epoch imaging technique used here requires short time delays between images, and so we preferentially select lenses with close image pairs (or high symmetry in the case of Q2237+0305). We have chosen close image pairs that exhibit strongly anomalous flux ratios for our analysis. Since a priori, anomalous flux ratios are driven by a smooth matter fraction \citep{sw02}, this selection biases our estimates towards larger smooth dark matter fractions. Thus our results cannot be used to make global statements regarding smooth dark matter in galaxies.  However eight out of ten close image systems display anomalous flux ratios \citep{pooley+07}, and so we expect the smooth matter percentage at those image positions to be high. This indicates that any bias in our analysis of individual systems would be relatively minor. In the future our analysis could be extended to a statistical sample through the use of monitoring data and the light curve technique discussed in \citet{k04}.

\section{Conclusions}
\label{sec:conclusions}
We have presented estimates of the smooth matter percentage in three lensing galaxies along the line of sight to the lensed images. We find a smooth matter percentage of $50^{+30}_{-40}$ in MG 0414+0534, $80^{+10}_{-10}$ in SDSS J0924+0219, and $\leq50$ in Q2237+0305, with 68 per cent confidence. In the two systems where the lensed images lie in the outer regions of the lensing galaxies (5 to 10 kpc from their centres), these measurements are inconsistent with zero smooth dark matter in the lensing galaxies. In Q2237+0305, where the lensed images lie in the central bulge of the lensing galaxy and so stars are expected to dominate the microlensing signal, our result is consistent with zero per cent smooth matter, as expected.

These measurements were obtained using a single-epoch imaging technique that is free from the need for long-term monitoring campaigns, which are required for detailed analysis of the lens mass profile. Our results also do not depend upon the typically unknown velocities of the stars in the lensing galaxy, and of the lensing galaxy and background source themselves, which enter into analyses of microlensing lightcurves. It is however only appropriate for systems with time delays between images of less than a day, so we can ensure we are observing the background source in the same state along each line of sight. It also does not provide us with any information on the slope of the dark matter profile; for that, stellar velocity dispersions in the lensing galaxy are required.

Nevertheless, our technique does provide an observationally inexpensive method for estimating the dark matter fraction in lensing galaxies at the location of lensed images. Gravitational lensing remains the only method for directly probing the dark matter content of galaxies.

\acknowledgments

NFB acknowledges the support of an Australian Postgraduate Award. DJEF acknowledges the support of a Magellan Fellowship from Astronomy Australia Limited. We are indebted to Joachim Wambsganss for the use of his inverse ray-shooting code. We thank the referee for comments which helped improve the final version of this manuscript.

{\it Facilities:} \facility{Magellan:Baade}, \facility{HST}, \facility{VLT:Kueyen}.

\end{document}